\documentclass[onecolumn,floatfix,superscriptaddress,a4paper,
               showpacs,showkeys,nofootinbib,preprint]{revtex4}
\usepackage{epsfig}
\usepackage{latexsym}
\usepackage{xspace}
\usepackage{hyperref}
\usepackage[latin2]{inputenc}
\usepackage{indentfirst}
\usepackage{enumerate}
\usepackage{color}

\usepackage{amsmath}
\usepackage{amssymb}
\usepackage[english]{babel}
\usepackage{url}

\newcommand{\mean}[1]{\langle #1 \rangle}

\newcommand{\Eq}[1]{Eq.~(\ref{#1})}

\newcommand{\eq}[1]{\begin{align} #1 \end{align}}

\begin{document}

\title{Open charm production in central Pb+Pb collisions \\ at the CERN SPS:
statistical model estimates
}

\author{R.~V. Poberezhnyuk}
\affiliation{Bogolyubov Institute for Theoretical Physics, Kiev, Ukraine}

\author{M. Gazdzicki}
 \affiliation{Goethe--University, Frankfurt, Germany}
 \affiliation{Jan Kochanowski University, Kielce, Poland}

\author{M. I. Gorenstein}
 \affiliation{Bogolyubov Institute for Theoretical Physics, Kiev, Ukraine}

 \affiliation{Frankfurt Institute for Advanced Studies, Frankfurt, Germany
\vspace{0.5cm}}

\begin{abstract}
Charm particle production in nucleus-nucleus collisions at the CERN SPS energies is considered
within a statistical approach. Namely, the Statistical Model of the Early Stage
is used to calculate mean multiplicity of charm particles in central Pb+Pb collisions.
A small number of produced charm particles
necessitates the use of the exact charm conservation law.
The model predicts a rapid increase of mean charm multiplicity as a function of collision energy.
The mean multiplicity calculated for central Pb+Pb collisions at the center of mass energy
per nucleon pair $\sqrt{s_{NN}}=17.3$~GeV exceeds significantly
the experimental upper limit. Thus, in order to describe open charm production
model parameters and/or assumptions should be revised.

\end{abstract}

\pacs{15.75.Ag, 12.40.Ee}

\keywords{Charm production, nucleus-nucleus collisions}

\maketitle

Model predictions concerning mean multiplicity of $c{\bar c}$-pairs produced
in central lead-lead collisions at the top CERN SPS energy, $E_{\rm lab}=158${\it A}~GeV,
differ significantly.
Perturbative-QCD calculations for proton-proton (p+p) interactions were done in Ref.~\cite{pQCD}.
Extrapolation of these results to central Pb+Pb collisions led to  estimate  $\langle N_{c{\bar c}}\rangle
\cong 0.17$  \cite{BM}.
The hadron resonance gas model with chemical freeze-out temperature $T\cong 170$~MeV gives
$\langle N_{c{\bar c}}\rangle =0.3\div 0.45$ \cite{GKSG}.
The ALCOR hadronization model \cite{levai} predicted $\langle N_{c{\bar c}} \rangle \cong 3.6$. Even
a larger yield, $\langle N_{c{\bar c}} \rangle \cong 8 $,
was expected in the Statistical Model of the Early
Stage (SMES) \cite{GG} which assumes statistical creation of charm quarks in a quark gluon
plasma (QGP). Thus, the model predictions vary almost by two orders of magnitude.

The model predictions for the system size dependence are also very different.
In the  perturbative-QCD inspired models, $\langle N_{c{\bar c}}\rangle $
is proportional to
$ N_p^{4/3}$, where $N_p$ is the number of
nucleon participants in Pb+Pb collisions. The dependence
$\langle N_{c{\bar c}} \rangle \sim N_p$  is expected in both, the SMES \cite{GG} and hadron-resonance gas model \cite{GKSG}.
A behavior  of  $\langle N_{c{\bar c}} \rangle \sim  N_p^{1.7}$ was suggested within the statistical coalescence model
\cite{GKSG}.

The NA49 Collaboration published \cite{NA49} an upper limit of 2.4 for
mean multiplicity of $D^0+ {\bar D}^0$ mesons produced in central Pb+Pb collisions at 158{\it A}~GeV.
This gives $\langle N_{c {\bar c}} \rangle<3.6$ if one assumes that, like in p+p interactions,
about one third of  ${\bar c}$ and $c$ quarks hadronizes into $D^0$ and ${\bar D}^0$ mesons.

The aim of the present paper is to
calculate collision energy dependence of open charm within the SMES
and discuss its dependence on model parameters related to charm production.
The
SMES describes the transition between confined and deconfined phases
of strongly interacting matter created in nucleus-nucleus collisions.
The model has
predicted several signals of the deconfinement phase transition \cite{GG,NA49,NA49a,NA49b,rustamov,review}, which were observed experimentally.

The SMES assumes that nucleons slow down and lose the fraction $\eta < 1$ of their initial energy in A+A central collisions.
They fly away carrying their baryonic and electric charges.
Therefore, the newly created matter with all conserved charges equal to zero is considered.
This matter is assumed to be statistically produced in longitudinally contracted fireball with volume:
\eq{\label{volume}
V=\frac{4 \pi r^3_0 N_p/3}{\sqrt{s_{NN}}/2m_N}~,~~
}
where $m_N~=~939~ \rm{MeV}$ is the nucleon mass, $\sqrt{s_{NN}}$ is the center of mass energy of the nucleon pair,
$N_p$ is the number of participant nucleons
from a single nucleus ($N_p=207$ for central Pb+Pb collisions is assumed). The $r_0$ parameter is taken to be 1.30~fm in order to fit the mean
baryon density in the nucleus, $\rho_0 = 0.11$~fm$^{-3}$.
The energy used for particle creation
(inelastic energy)  is assumed to be:
\eq{\label{E}
E~=~\eta \,(\sqrt{s_{NN}}~-~2m_N)\,N_p~,
}
where
parameter $\eta~=~0.67$ \cite{GG}.

Since the system of newly created particles has all conserved charges equal to zero,
the pressure $p$ and energy density $\varepsilon=E/V$ are assumed
to be functions of temperature $T$ only. These functions in the confined (W-phase)  and deconfined (Q-phase) phases are equal
to the (almost) ideal gas ones,
where massless
non-strange hadron and quark-gluon degrees of freedom have the degeneracy factors $g_W~=~16$ and $g_Q~=~37$.
Strange constituents are considered as massive with $m^s_Q\cong 200$~MeV and $g^s_Q=12$
in the quark-gluon phase, and
$m^s_W = 500$~MeV, $g^s_W~=~14$ in the confined phase.
For Q-phase
the bag model equation of state is used~\cite{bag}:
$p_Q=p_Q^{\rm id}-B$ and $\varepsilon_Q=\varepsilon_Q+B$.
Thus, the bag constant is added to the ideal gas of quarks and gluons.
It is chosen to fix the value of the phase transition temperature $T_c = 200$~MeV.
Note that the lattice QCD data suggests the crossover transition temperature $T_c = 150-170$~MeV.
However, in the present paper we keep the value  $T_c=200$~MeV, as used in the original SMES formulation \cite{GG}.
A revision of the SMES with self-consistent changes of  all model parameters is outside of the scope of the present paper.

The entropy densities in the pure phases ($i=$W, Q) read:
\eq{\label{s}
s_i(T)~=~\frac{p_i(T)~+~\varepsilon_i (T)}{T}~.
}
%
%

The energy and entropy densities in the mixed phase are ($0<\xi<1$):
\eq{\label{e-mix}
&\varepsilon_{\rm mix}(T_c)~=~\xi\,\varepsilon_Q (T_c)~+~(1-\xi)\,\varepsilon_W (T_c)~,\\
%
%
&s_{\rm mix}(T_c)~=~\xi\,s_Q (T_c)~+~(1-\xi)\,s_W (T_c)~.\label{s-mix}
}
%

The mixed phase starts at collision energy $\sqrt{s_{NN,1}}$ and ends at $\sqrt{s_{NN,2}}$:
\eq{\label{s1s2}
\sqrt{s_{NN,1}}~=~7.42~{\rm GeV},~~~~\sqrt{s_{NN,2}}~=~10.83~{\rm GeV}~.
}
%

We introduce now charm degrees of freedom assuming that
mean multiplicity of charm carriers is small ($\lesssim 1$). This assumption has two consequences:

(i) one can neglect the contribution of charm degrees of freedom to  energy density and  pressure of the system
(thus, the phase transition location
remains unchanged);

(ii) one has to consider
the canonical ensemble (CE) for charm particles that assures
an equal number of charm and
anti-charm charges
in each microscopic state
of the system.

The CE was used previously to calculate mean multiplicity of strange particles in p+p interaction \cite{Poberezhnyuk:2015wea}.
Within the SMES, similarly to the strangeness case, the CE formulation
for  charm leads to a suppression of  charm yield with respect to the grand canonical ensemble (GCE) yield by a factor
equal to the ratio of the Bessel
functions $I_1$ and $I_0$:
\eq{\label{nc-CE}
n_{W,Q}^{c ({\rm CE})}(T,V)~
 =~n^{c }_{W,Q}(T) ~\frac{I_1\left[V n^{c}_{W,Q}(T)\right]}{I_0\left[V n^{c}_{W,Q}(T) \right]}~,
 %
 %
}
where the number density of the sum of charm and anti-charm particles in the GCE for pure phases
can be calculated as
\eq{\label{nW-c}
 n^{c}_{W,Q}(T)~=~\frac{g^{c }_{W,Q}}{2 \pi^2} \int^{\infty}_0 dk\,k^2 \exp\left[-~\frac{\sqrt{k^2+(m^c_{W,Q})^2}}{T}~\right]~,
%
}
with $m_W^c\cong 1.9~ \rm{GeV}$ being a $D$-meson mass, $m_{Q}^c \cong 1.3~\rm{GeV}$ being a charm quark mass. The degeneracy factor for the (anti-)charm
quarks is $g^{c}_{Q}=12$, whereas the degeneracy factor for the (anti-)charm particles in the confined phase, $g^{c}_W$, is a free parameter.
In the mixed phase Eq.~(\ref{nc-CE}) should be
replaced by
\eq{\label{nc-CE-mix}
n^{c({\rm CE})}_{\rm mix}(T,V,\xi)~=~X~\frac{I_1[X]}{I_0[X]}~,
}
where
\eq{\label{X}
X~=~X(T,V,\xi)~ =~\xi\,V\, n^{c}_{Q}(T)~+~(1-\xi)\,V\, n^{c}_{W}(T)~
}
is the mean number of charm and anti-charm particles
in the mixed phase calculated within the GCE.
At each $\sqrt{s_{NN}}$ one calculates $V$ and $\varepsilon =E/V$
according to Eqs.~(\ref{volume}) and (\ref{E}), and then mean multiplicity
of $c {\bar c}$-pairs is calculated as
\eq{\label{Ncc}
\langle N_{c {\bar c}}\rangle ~=~\frac{1}{2}~n^{c\,({\rm CE})}\,V~,
}
%
%
where $n^{c (CE)}$ is given by \Eq{nc-CE} in the pure phases or by \Eq{nc-CE-mix} in the
mixed phase.

In the mixed phase, the temperature
$T$ and the parameter $\xi$ are obtained by solving the equations:
\eq{\label{e-CE-mix}
 \xi~=~\dfrac{\varepsilon(\sqrt{s_{NN}})~-~\varepsilon_W(T_c)}{\varepsilon_Q(T_c)~-~\varepsilon_W(T_c)}~,
~~~~~ p_Q(T_c)~=~p_W(T_c)~.
%
}

\begin{figure}[!htb]
\includegraphics[width=0.49\textwidth]{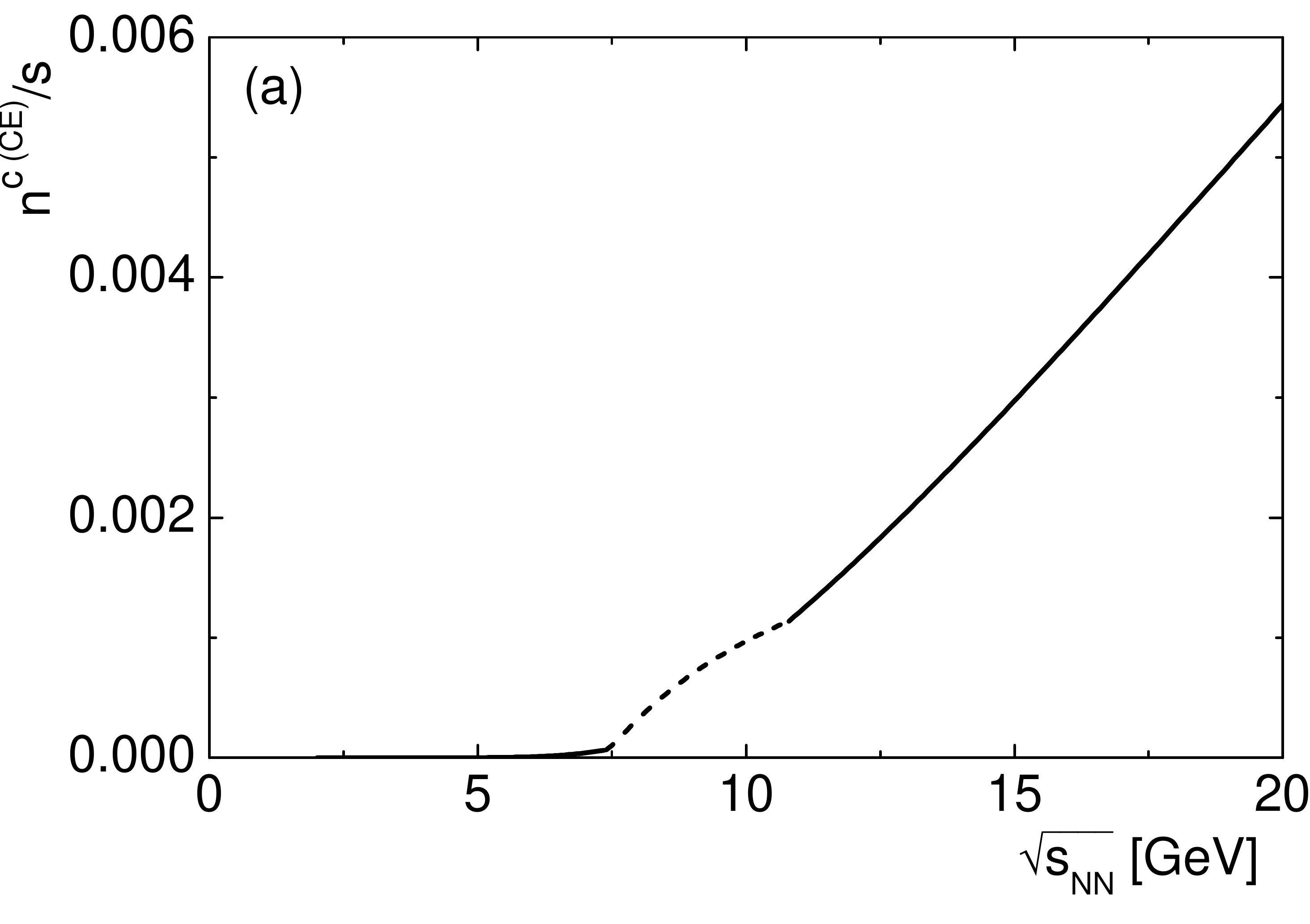}
\includegraphics[width=0.49\textwidth]{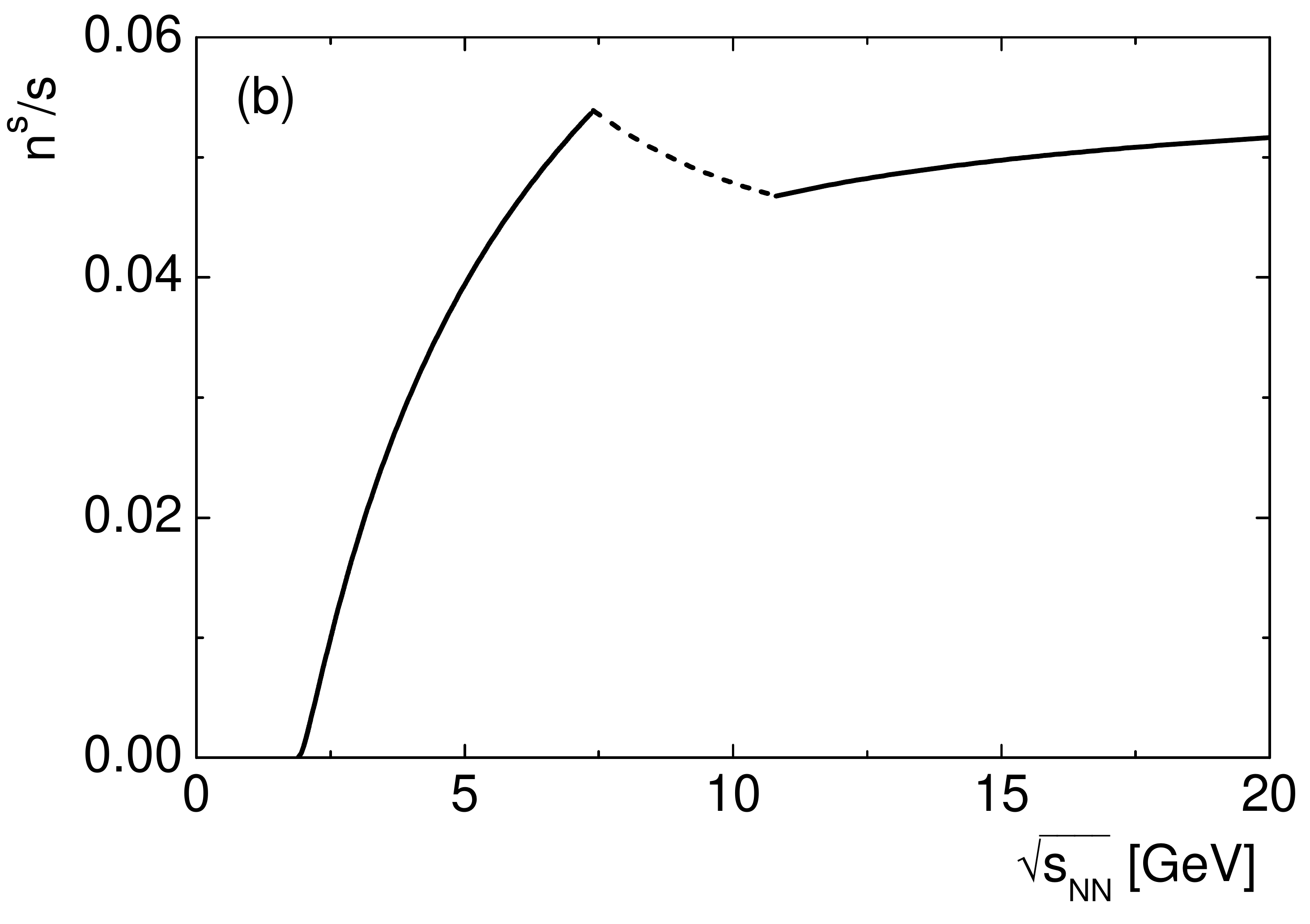}
\caption{
Collision energy dependence of the charm to entropy ratio  ($a$) and of the strangeness to entropy ratio  ($b$)
calculated within the SMES for central Pb+Pb collisions.
Dashed lines denote the mixed phase region.
\label{fig-horn}
}

\end{figure}
The charm to entropy ratio calculated for central Pb+Pb
collisions for $g_W^c=10$
is plotted in Fig.~\ref{fig-horn}~$(a)$ as a function
of collision energy.
The strangeness to entropy ratio
is plotted in Fig.~\ref{fig-horn}~$(b)$ for a comparison.
While the behavior of strangeness to entropy ratio
exhibits the {\it horn} structure  \cite{GG}, the charm to entropy
ratio is a monotonous function of collision energy.
\begin{figure}[!htb]
\includegraphics[width=0.49\textwidth]{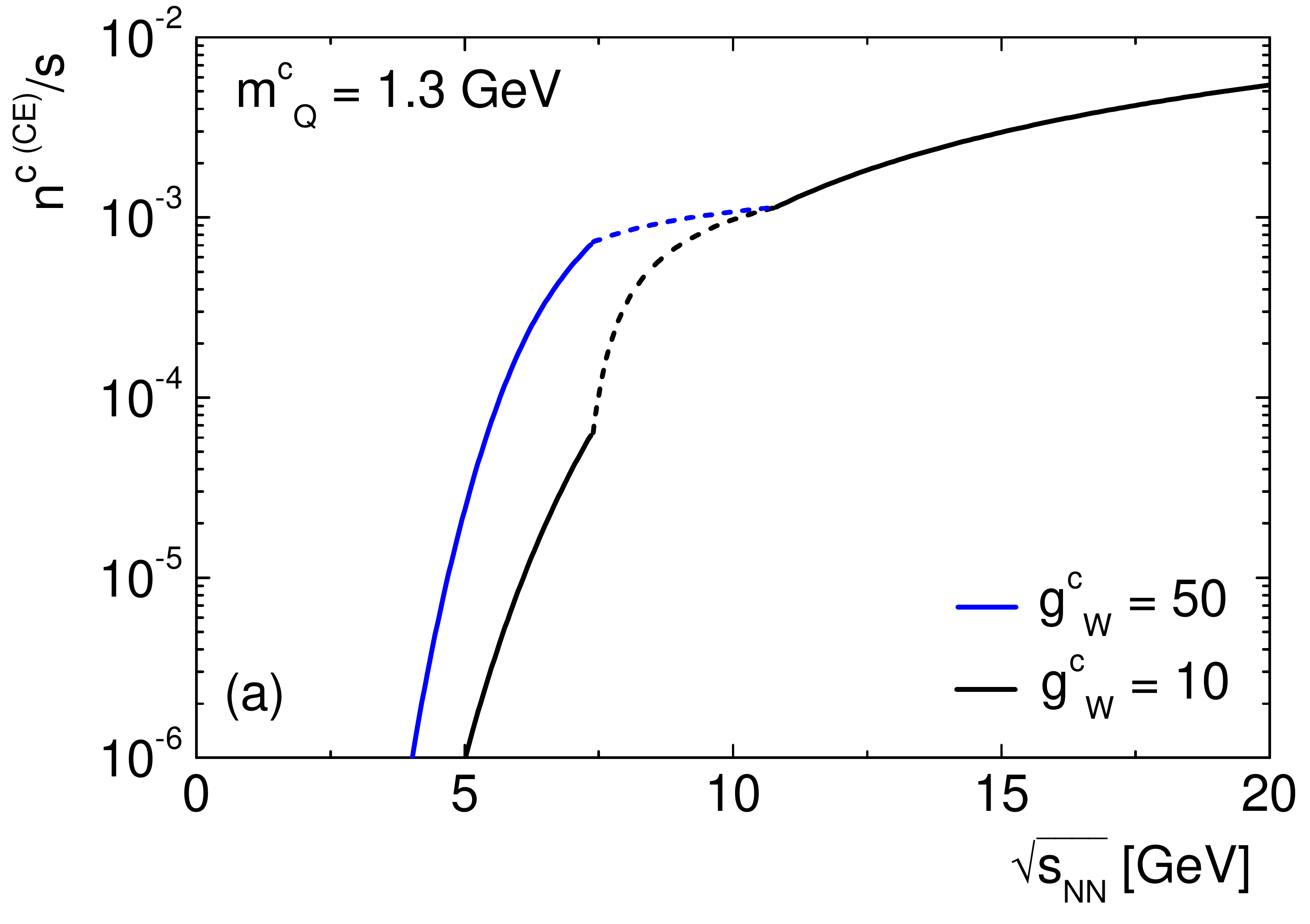}
\includegraphics[width=0.49\textwidth]{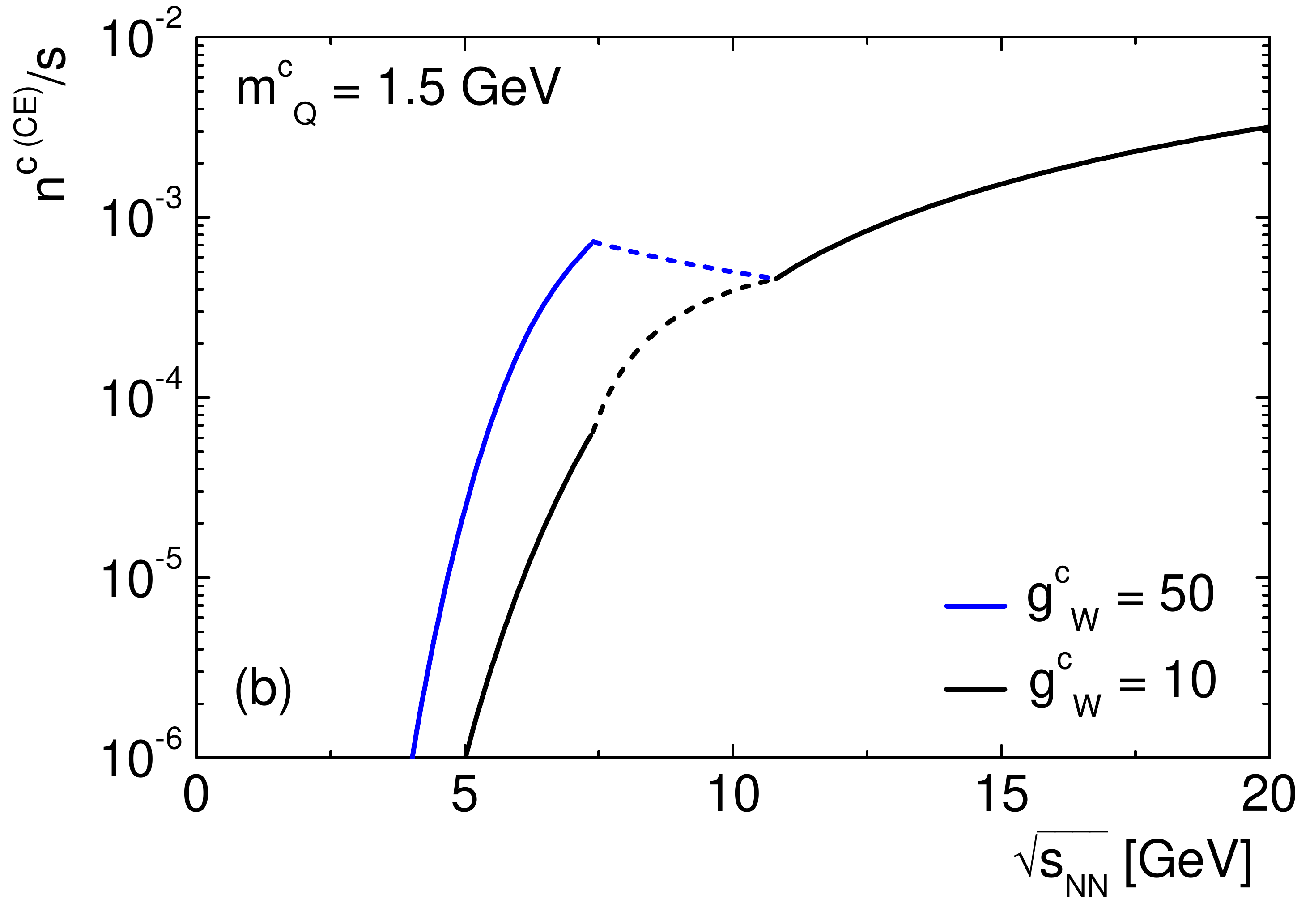}
\includegraphics[width=0.49\textwidth]{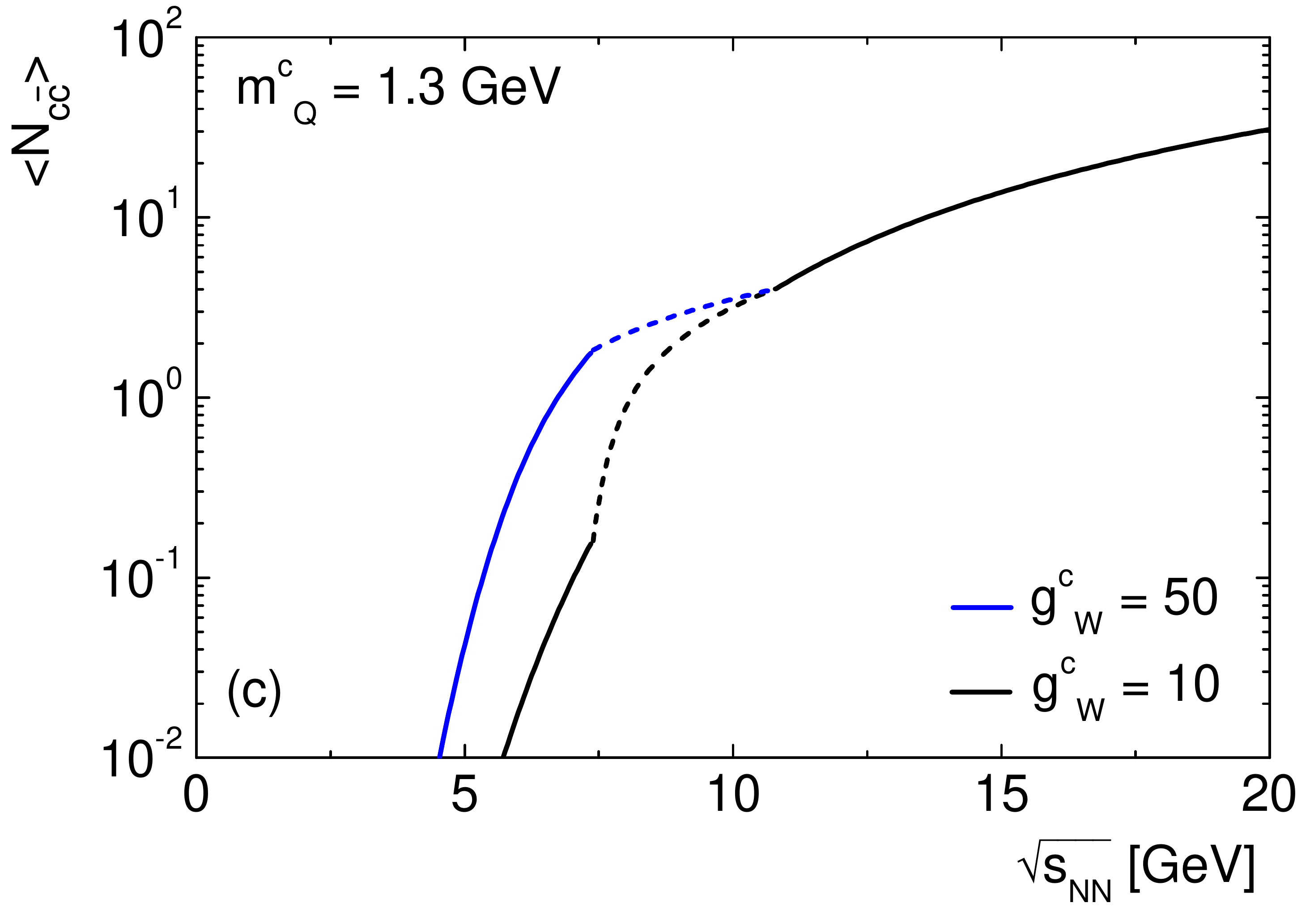}
\includegraphics[width=0.49\textwidth]{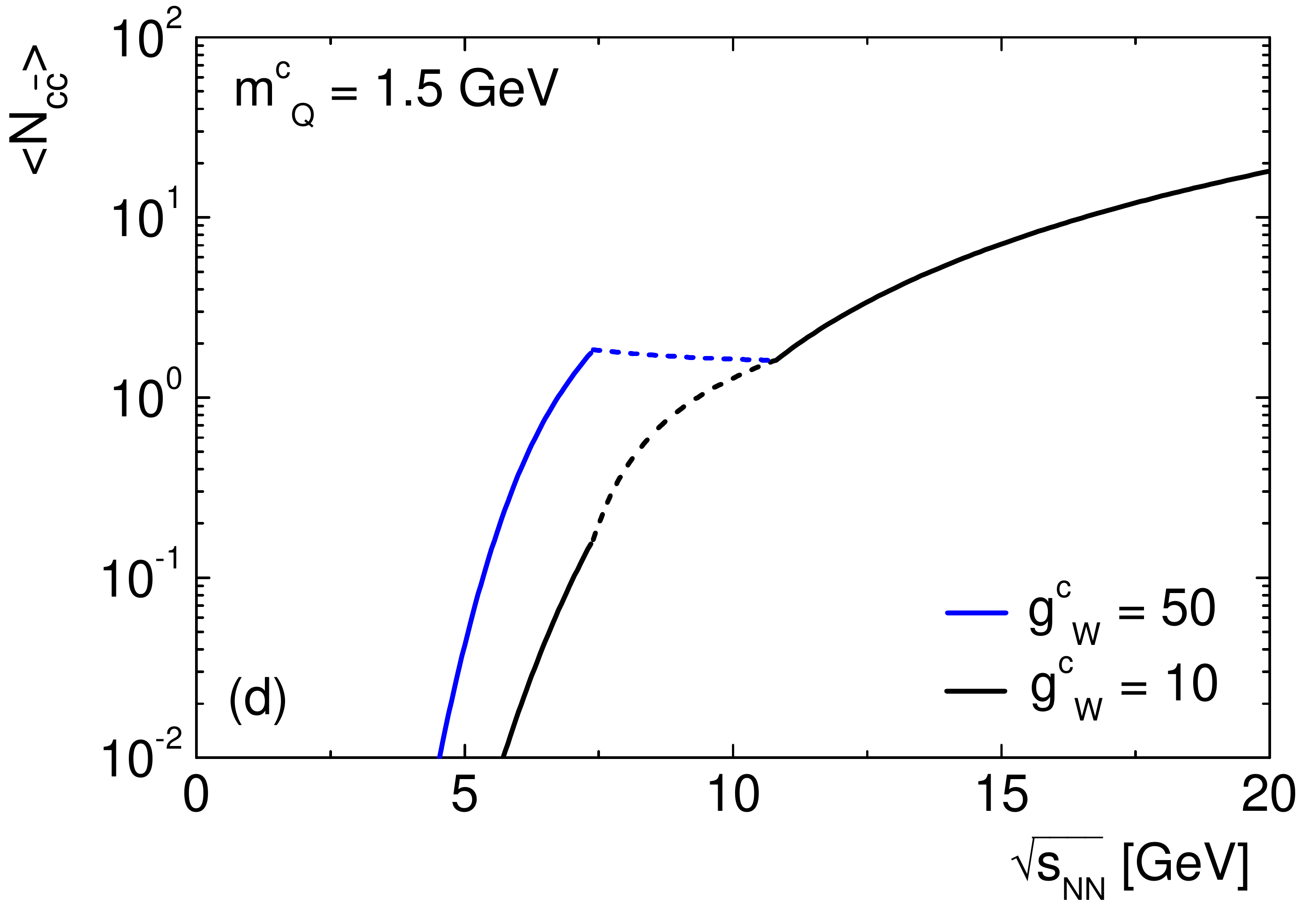}
\caption{
Collision energy dependence of the charm to entropy ratio ($a,b$) and a mean  multiplicity of $c\bar{c}$-pairs ($c,d$)
calculated within the SMES for central Pb+Pb collisions  with the mass of the charm
quark 1.3 GeV ($left$) and 1.5 GeV ($right$) and charm particle degeneracy factor $g^{c }_W = 10$ (lower lines) and $g^{c}_W = 50$ (upper lines).
\label{fig-horn2}
}
\end{figure}

Figure \ref{fig-horn2} shows a collision energy dependence of results on open charm for $m_Q^c=1.3$~GeV and 1.5~GeV,
and $g_W^c =10$ and 50.
As seen from Fig.~\ref{fig-horn2}, the horn structure is absent in the charm to entropy ratio for $g^{c}_W = 10$ and $m^c_Q=1.3$~GeV,
but appears for large (unphysical) values of $g^{c}_W = 50$ and $m^c_Q=1.5$~GeV.

\begin{figure}[!htb]
\includegraphics[width=0.59\textwidth]{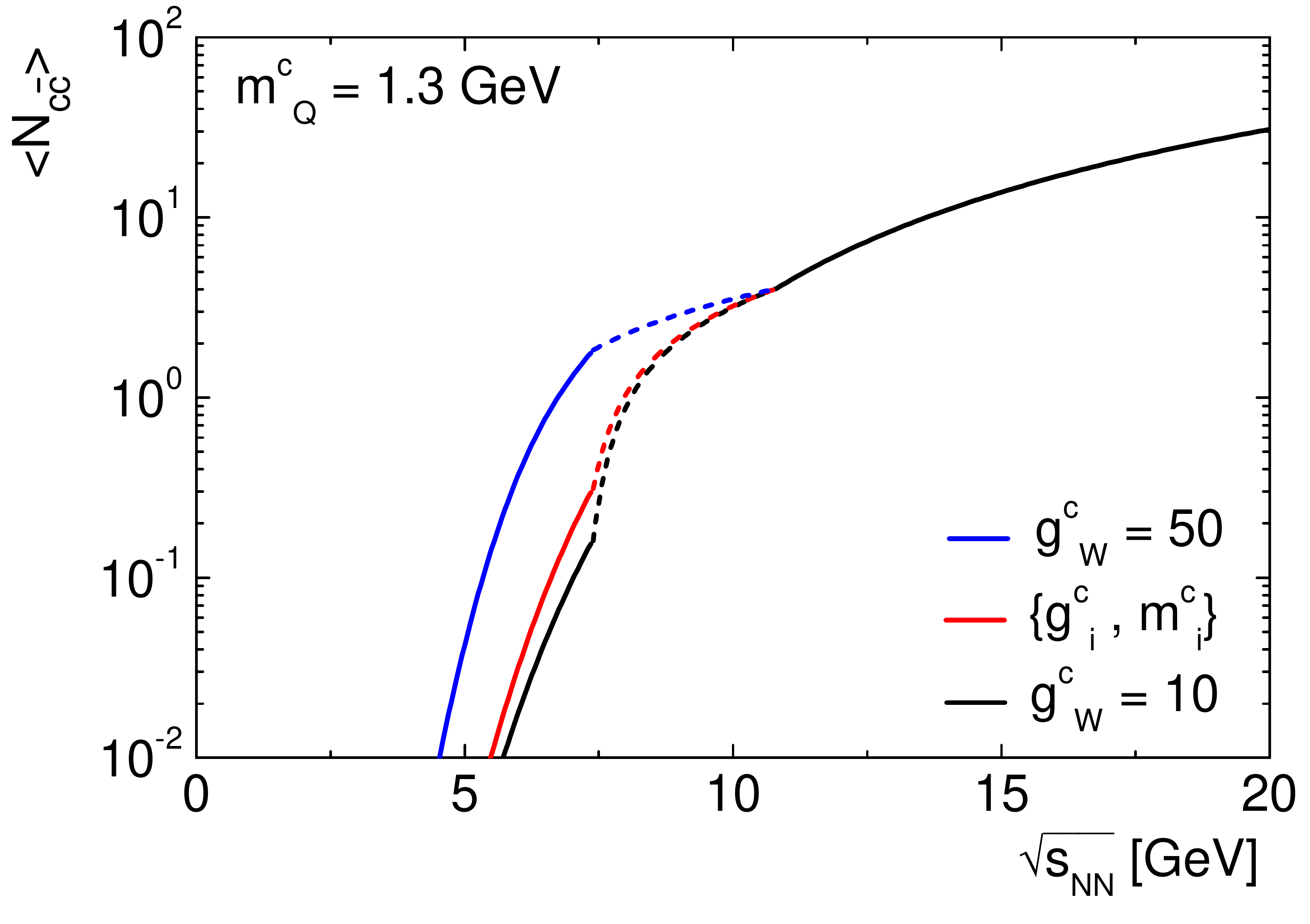}
\caption{
Collision energy dependence of a mean  multiplicity of $c\bar{c}$-pairs
calculated within the SMES for central Pb+Pb collisions  with the mass of the charm
quark 1.3 GeV. Black and blue lines correspond to the assumption of $D$-meson mass $m^{c }_W=1.9$~ GeV and 
effective degeneracy factors $g^{c }_W = 10$ and $g^{c }_W = 50$, respectively. Red line is  obtained
by accounting contributions (\ref{nW-c1}) from all non-strange $D$-mesons with corresponding masses and degeneracy factors $\{m^c_i,g^c_i\}$.
\label{fig-horn3}
}
\end{figure}

To  estimate the proper value of parameter $g^c_W$ let us consider charm and anti-charm in the white phase as a sum of all $D$-meson states,
\eq{\label{nW-c1}
 n^{c}_{W}(T)~=~\sum_{i}\frac{g^{c }_{i}}{2 \pi^2} \int^{\infty}_0 dk\,k^2 \exp\left[-~\frac{\sqrt{k^2+(m^c_{i})^2}}{T}~\right]~.
}
In sum (\ref{nW-c1}) we include 7 non-strange charm mesons from $D^0$ ($i=1$) with $m^c_1=1.86$~GeV and $g^c_1=2$ 
and $D^{\pm}$ ($i=2$) with $m^c_2=1.87$~GeV and $g^c_2=2$  up to $D_2^{*0}$ and $D_2^{*\pm}$ ($i=6,7$), both  with $m^c_{6,7}=2.46$~GeV and $g^c_{6,7}=10$.
In Fig.~\ref{fig-horn3} the results of Eq.~(\ref{nW-c1}) for energy dependence of $\mean{N_{c\bar{c}}}$ in the CE  are  compared with those 
obtained from  \Eq{nW-c} for $n^c_W$ with mass $m^c_W=1.9$~GeV and effective degeneracy factors $g^c_W=10$ and $g^c_W=50$.
The results for the full spectrum of charm mesons are higher (by a factor 1.9 in the beginning of the mixed phase) than the results for $g^c_W=10$, considered as lower limit, whereas they are significantly lower (by a factor 5.8 in the beginning of the mixed phase) than the results for $g^c_W=50$.

In summary, collision energy dependence of mean number of $c{\bar c}$-pairs in
central Pb+Pb collisions was calculated within the Statistical Model of the Early Stage.
Mean charm multiplicity and its ratio to entropy exhibit a rapid growth as functions
of collision energy in the considered energy region. In central Pb+Pb collisions at
$\sqrt{s_{NN}}=17.3$~GeV, $\mean{N_{c {\bar c}}}\cong 20$ for $m_Q^c=1.3$~GeV and
$\mean{N_{c {\bar c}}}\cong 8$ for $m_Q^c=1.5$~GeV. These values are significantly
larger than the experimental bound, $\mean{N_{c {\bar c}}}\cong 3.6$, reported in Ref.~\cite{NA49}.
The SMES predictions are sensitive to assumed value of charm quark mass and charm degeneracy
factor in the confined matter. But even for extreme values of these parameters the SMES
predictions disagree with the experimental data. Thus, a quantitative description of charm production
within SMES requires a revision of parameters and/or assumptions of the model.

\begin{acknowledgements}
The work of M.I.G. is supported by the Goal-Oriented Program of the National Academy of Science of Ukraine,
by the European Organization for Nuclear Research (CERN), Grant CO-1-3-2016, and by the Program of
Fundamental Research of the Department of Physics and Astronomy of the National Academy of Science of Ukraine.
\end{acknowledgements}


\begin{thebibliography}{105}


\bibitem{pQCD}
R. V. Gavai {\it et al.} Int. J. Mod. Phys. A {\bf 10} 2999 (1995).

\bibitem{BM}
P. Braun-Munzinger and J. Stachel, Phys. Lett. B {\bf 490},
196 (2000).

\bibitem{GKSG}  M. I. Gorenstein, A.~P.~Kostyuk, H.~Stoecker, and W.~Greiner,
Phys. Lett.
B {\bf 509}, 277 (2001).

\bibitem{levai}
P. Levai, T. S. Biro, P. Csizmadia, T. Csorgo, and J. Zimanyi J. Phys. G {\bf 27}, 703 (2001).





\bibitem{GG} M. Gazdzicki and M.~I.~Gorenstein, Acta Phys. Pol. {\bf B30}, 2705 (1999).

\bibitem{NA49}
C. Alt et al., [NA49 Collaboration] Phys. Rev. C {\bf 73}, 034910 (2006).

\bibitem{NA49a}
  C.~Alt {\it et al.}  [NA49 Collaboration],
  Phys.\ Rev.\  C {\bf 77}, 024903 (2008).


 \bibitem{NA49b}
 S.~V.~Afanasiev {\it et al.}  [NA49 Collaboration],
  Phys.\ Rev.\ C {\bf 66}, 054902 (2002).


\bibitem{rustamov} A. Rustamov, Central Eur. J. Phys. {\bf 10}, 1267 (2012).


\bibitem{review}
  M.~Gazdzicki, M.~I.~Gorenstein, and P.~Seyboth,
  Acta Phys.\ Polon.\ B {\bf 42}, 307 (2011); \\
  M.~Gazdzicki, M.~I.~Gorenstein, and P.~Seyboth,
  Int. J. Mod. Phys. E {\bf 23}, 1430008 (2014).



\bibitem{bag} J. Baacke, Acta Phys. Polon. B
{\bf 8}, 625 (1977); \\
 E. V. Shuryak, Phys. Rept. {\bf 61}, 71 (1980); \\
J. Cleymans, R. V. Gavai and E. Suhonen, Phys. Rept. {\bf 130}, 217 (1986).

\bibitem{Poberezhnyuk:2015wea}
  R.~V.~Poberezhnyuk, M.~Gazdzicki and M.~I.~Gorenstein,
  Acta Phys.\ Polon.\ B {\bf 46},
  1991 (2015).















 \end{thebibliography}
\end{document}